\documentclass[aps,prd,superscriptaddress,twoside,twocolumn,nofootinbib,10pt,
showpacs,floatfix]{revtex4-1}
\usepackage{amsmath,amssymb}
\usepackage{graphicx,bm}
\usepackage{slashed}
\usepackage{dcolumn}
\usepackage{epstopdf}
\usepackage{ulem} 
\usepackage[usenames]{color}
\allowdisplaybreaks
\newcommand{\dd}{{\rm d}}              
           
\newcommand{\vc}[1]{\textbf{\emph #1}}
\newcommand{\mnras}{MNRAS~}              \newcommand{\na}{New Astronomy~}
\newcommand{\araa}{ARA\&A~}              \newcommand{\aap}{A\&A~}
\begin{document}
\title{Data-based optimal bandwidth for kernel density estimation of statistical samples}
\author{Zhen-Wei Li}
\affiliation{Center for Theoretical Physics and College of Physics, Jilin University, Changchun,
130012, China}
\affiliation{Changchun Observatory, National Astronomical Observatories, CAS, Changchun 130117, China}
\author{Ping He}
\email{hep@jlu.edu.cn}
\affiliation{Center for Theoretical Physics and College of Physics, Jilin University, Changchun,
130012, China}
\affiliation{Center for High Energy Physics, Peking University, Beijing 100871, China}
\date{\today}
\begin{abstract}
\noindent {\bf Abstract} It's a common practice to evaluate probability density function or matter spatial density function from statistical samples. Kernel density estimation is a frequently used method, but to select an optimal bandwidth of kernel estimation, which is completely based on data samples, is a long-term issue that has not been well settled so far. There exist analytic formulae of optimal kernel bandwidth, but they cannot be applied directly to data samples, since they depend on the unknown underlying density functions from which the samples are drawn. In this work, we devise an approach to pick out the totally data-based optimal bandwidth. First, we derive correction formulae for the analytic formulae of optimal bandwidth to compute the roughness of the sample's density function. Then substitute the correction formulae into the analytic formulae for optimal bandwidth, and through iteration we obtain the sample's optimal bandwidth. Compared with analytic formulae, our approach gives very good results, with relative differences from the analytic formulae being only $2 \sim 3\%$ for sample size larger than $10^4$. This approach can also be generalized easily to cases of variable kernel estimations.

{\vskip 10pt}

\noindent {\bf Key words:} numerical methods, kernel density estimation, optimal bandwidth, large-scale structure of Universe
\end{abstract}

\maketitle
\section{Introduction}
\label{sec:intro}

In statistics and many other scientific disciplines such as modern cosmology, it is a common practice to estimate probability density function from statistical data, or matter density function in a given spatial region from observational or simulation samples. In particular, $N$-body simulations are extensively used to investigate the distribution, formation and evolution of large-scale nonlinear structures of the Universe, such as galaxies, galaxy clusters, filaments and inter-galactic gas, and so forth \cite{efs85,bert98,Springel01}. For example, density profiles of dark matter halos are frequently investigated, and these density profiles of dark matter halos are usually considered as universal \cite{nfw95,nfw96,nfw97,moore99,einasto65,navarro10} and widely used in various cosmological studies. Hence, it is of great significance to evaluate matter densities of nonlinear structures from simulation data samples.

There are many excellent methods to evaluate matter density from cosmological simulation samples, such as Delaunay Tessellation Field Estimator \cite{Schaap2000}, or the dark-matter sheet method \cite{Abel2012,Shandarin2012}. In statistics or cosmology, however, the widely used ones are still histogram method and kernel estimation method \cite{silverman86}. They are parametric methods, i.e. they depend on the binwidth for histogram or on the bandwidth for kernel method. So how to make the optimal binwidth or bandwidth is crucial for density estimation from statistical samples, e.g. Refs.~\cite{scott92,gentle09}.

Kernel estimation is believed to be superior to histogram method \cite{jones96}, since if the kernel is a derivable function, then the estimated density function is also derivable. Hence, in this work, we only consider the kernel estimation method.

A non-negative real-valued function $K_1(u)$ is called a kernel function, if satisfying the following two conditions:
\begin{equation}
\label{eq:kernel}
\int_{-\infty}^{+\infty}K_1(u)\dd u =1; {\hskip 20pt}  K_1(u)=K_1(-u).
\end{equation}
The density function $f(x)$ of a univariate statistical sample, $x_i$, with $i$ running from 1 to the sample size $N_p$, can be estimated by using the kernel function as:
\begin{equation}
\label{eq:funkh}
\hat{f}_h(x) = \frac{1}{N_p h}\sum\limits_{i=1}^{N_p}K_1(\frac{x-x_i}{h}),
\end{equation}
in which $h$ is the bandwidth of the kernel.

As aforementioned, for a proper density estimation, it is crucial to select an appropriate bandwidth for the kernel function. A small bandwidth $h$ makes the estimated $\hat{f}_h$ too rough and produces spurious features, while a large bandwidth over-smoothes the data sample, so that useful features are lost \cite{jones96}. Hence, the most important task for kernel estimation is to choose the optimal bandwidth $h$ for the data sample.

It is possible to select an optimal bandwidth of kernel method by computing the entropy $S[\hat{f}_h] = -\int \hat{f}_h(x) \ln \hat{f}_h(x) \dd x$ with the estimated density $\hat{f}_h(x)$ \cite{wolpert13,hep14}. Yet, it is very time-consuming to pick out the optimal bandwidth by evaluating the entropy, and hence it is not a proper approach to select the optimal bandwidth.

There are indeed analytical formulae of optimal bandwidth for kernel estimation, which are obtained by minimizing the so-called {\it asymptotic mean integrated squared error} (AMISE) \cite{silverman86, jones96}. The optimal bandwidth formulae, however, cannot be applied directly to data samples, since they depend on the underlying density function $f(x)$, from which the samples are drawn, but $f$ is usually not known. So some data-based methods of selecting the optimal bandwidth have been developed, such as the `cross validation' methods \cite{Rudemo1982,Bowman1984,Hall1992}, and `plug-in' methods \cite{Sheather1991, Botev2010}. For a comprehensive review about selecting the optimal bandwidth of kernel estimation, see Ref.~\cite{jones96}.

In this work, based on the analytic formulae of optimal bandwidth, we devise a practical method on how to select an optimal bandwidth from data samples without the knowledge of the underlying density function. In Section~\ref{sec:opth}, we provide the analytic optimal bandwidth of kernel estimation for univariate samples. We introduce the method in Section~\ref{sec:method}, and present the results, the generalization to three-dimensional (3D) cases, and a cosmological application in Section~\ref{sec:resapp}. In Section~\ref{sec:summy}, we give summary and conclusions of our results.

\begin{figure}
\centerline{\includegraphics[width=1.0\columnwidth]{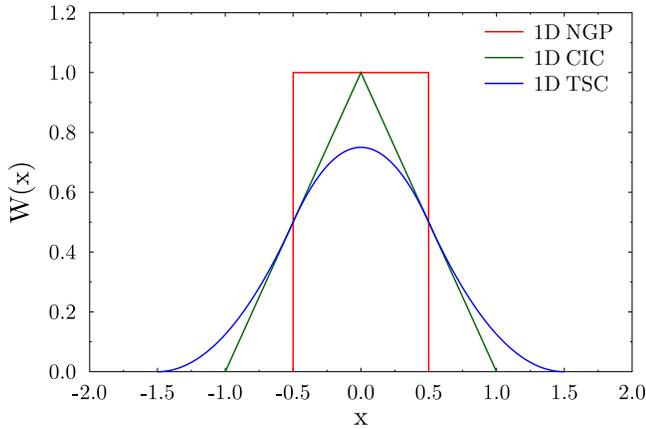}}
\caption{The one-dimensional mass assignment function of NGP, CIC and TSC scheme. In this work, we do not use them to assign particle mass to mesh points, but use them as kernel functions for density estimation of univariate samples.}
\label{fig:wx}
\end{figure}

\section{Analytic Optimal Bandwidth of Kernel Estimation for Univariate Samples}
\label{sec:opth}

For a univariate statistical sample, a common way of measuring the error in the estimation process is the mean integrated squared error (MISE),
\begin{equation}
\label{eq:mise}
{\rm MISE}(h) = E \int(\hat{f}_h(x)-f(x))^2\dd x,
\end{equation}
where $E$ denotes the expected value with respect to that sample. MISE is asymptotically (as $N_p\rightarrow \infty$) approximated by the AMISE \cite{silverman86, jones96},
\begin{eqnarray}
\label{eq:amise}
{\rm AMISE}(h) = \frac{R_1(K_1)}{h N_p} + h^4R_1(f'')\left(\int x^2 \frac{K_1(x)}{2}\dd x \right)^2,
\end{eqnarray}
in which $f''$ is the second derivative of $f(x)$ with respect to $x$, and the functional $R_1(g)$ is defined as
\begin{equation}
\label{eq:rg1}
R_1(g)\equiv\int_{-\infty}^{+\infty}g(x)^2\dd x.
\end{equation}
By minimizing the AMISE, i.e. $\dd ({\rm AMISE})/\dd h =0$, one obtains the optimal bandwidth $h_{\rm opt}$ for univariate statistical samples as
\begin{equation}
\label{eq:hopt1}
h_{\rm opt}=\left[\frac{R_1(K_1)}{R_1(f'')(\int x^2 K_1(x)\dd x)^2}\right]^{1/5} N_p^{-1/5}.
\end{equation}

Eq.~(\ref{eq:hopt1}) cannot be directly applied to statistical samples, since it depends on the unknown density function $f$ and its second derivative $f''$.

\begin{figure}
\centerline{\includegraphics[width=1.0\columnwidth]{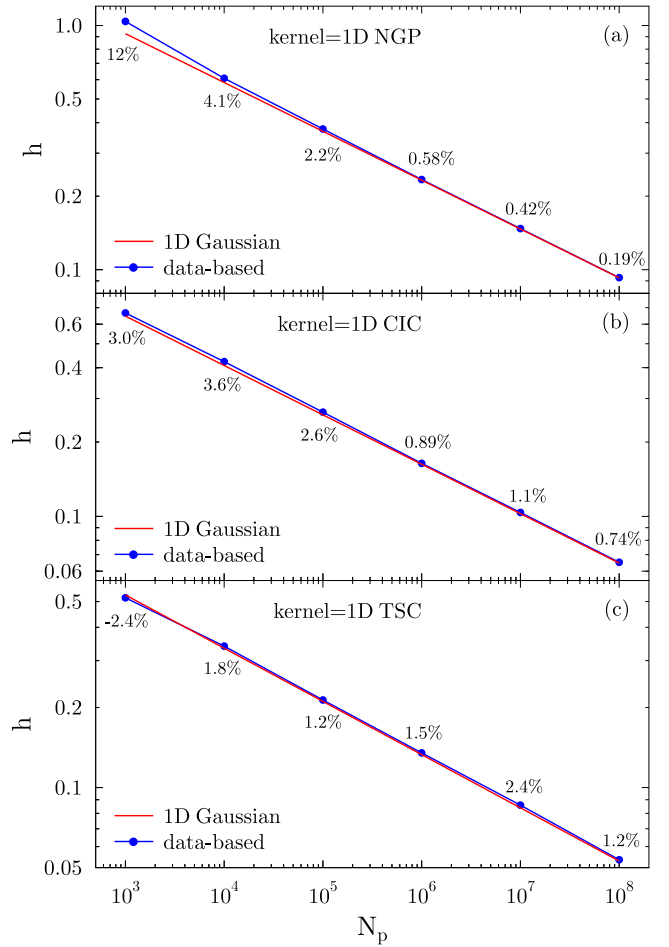}}
\caption{The optimal bandwidth $h$ vs. the sample size $N_p$ for univariate samples. The red solid lines indicate the analytic bandwidth of Eq.~(\ref{eq:hopt1}). While the blue solid lines indicate results of our approach, which is purely data-based. The percentage differences between the two are explicitly shown in the figure. The density function is the 1D Gaussian function of Eq.~(\ref{eq:1dgauss}). Panels (a), (b) and (c) are for kernel functions of 1D NGP, 1D CIC and 1D TSC, respectively.}
\label{fig:hopt-1}
\end{figure}

\begin{figure}
\centerline{\includegraphics[width=1.0\columnwidth]{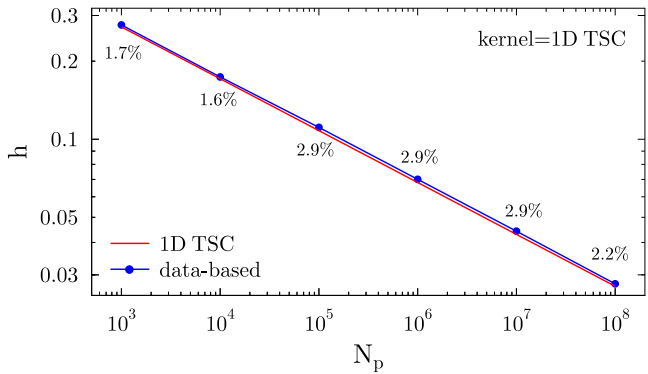}}
\caption{The same as Fig.~\ref{fig:hopt-1}, but the density function for the experimental sample is the 1D TSC function of Eq.~(\ref{eq:tsc}). The kernel function we adopted is also 1D TSC function.}
\label{fig:hopt-2}
\end{figure}

\section{Method}
\label{sec:method}

In $N$-body simulation codes such as those based on particle-mesh scheme \cite{hockney81}, particle masses are assigned to a set of pre-defined mesh points according to some mass assignment scheme. The usual mass assignment approaches are Nearest Grid Point (NGP), Cloud In Cell (CIC) and Triangular Shaped Cloud (TSC) schemes. The NGP function is
\begin{eqnarray}
\label{eq:ngp}
W_{\rm NGP}(x)=\left\{\begin{matrix}
 & 1, & |x| \leq 1/2, \\
 & 0,           & {\rm otherwise},
\end{matrix}\right.
\end{eqnarray}
the CIC function is
\begin{eqnarray}
\label{eq:cic}
W_{\rm CIC}(x)=\left\{\begin{matrix}
 & 1 - |x|, & |x| \leq 1, \\
 & 0,           & {\rm otherwise},
\end{matrix}\right.
\end{eqnarray}
and the TSC function is
\begin{eqnarray}
\label{eq:tsc}
W_{\rm TSC}(x)=\left\{\begin{matrix}
 & \frac{3}{4} - x^2, & |x| \leq 1/2, \\
 & \frac{1}{2}(\frac{3}{2}-|x|)^2, & 1/2 \leq |x| \leq 3/2, \\
 & 0,           & {\rm otherwise}.
\end{matrix}\right.
\end{eqnarray}
These functions are shown in Fig.~\ref{fig:wx}. It is interesting to notice that these mass assignment functions can also be used as kernel functions, since they satisfy the two conditions of Eq.~(\ref{eq:kernel}).

In order to demonstrate our approach of picking out the optimal bandwidth $h$ for univariate samples, we generate some experimental samples with known density functions, from which we draw $N_p$ random data points with Monte Carlo technique.

For a bandwidth $h$, the roughness $R_1(f'')$ in Eq.~(\ref{eq:hopt1}) is approximated by
\begin{equation}
\label{eq:chopt1}
R_1(f'') \approx R_1(\hat{f}''_h) - \frac{6}{w h^5 N_p},
\end{equation}
in which $w$ is the width of the kernel function. For a kernel function of finite extension, We define the non-zero range of a kernel function as the width of the kernel, and hence $w=1$, $2$ and $3$ for NGP, CIC and TSC, respectively. We give the derivation of Eq.~(\ref{eq:chopt1}) in Appendix-\ref{sec:apda}.

Our approach to pick out the optimal bandwidth $h$ is briefly described as follows.
\begin{enumerate}

\item With $W_{\rm NGP}(x)$, $W_{\rm CIC}(x)$ or $W_{\rm TSC}(x)$ as one-dimensional (1D) kernel functions, we construct the density estimator $\hat{f}_h(x)$ of univariate samples by Eq.~(\ref{eq:funkh}).

\item With this estimated $\hat{f}_h(x)$, we calculate $R_1(f'')$ by Eq.~(\ref{eq:chopt1}), and put it into Eq.~(\ref{eq:hopt1}) to update the bandwidth $h$.

\item Iterate step 1 to 2 until the relative difference of $h$ between two adjacent iterations is smaller than a tolerance parameter, say $\leq 0.1\%$.

\item Since an extremely small $h$ will make $R_1(f'')$ in Eq.~(\ref{eq:chopt1}) negative and hence meaningless, we choose a slightly larger $h$ as the initial value of the bandwidth.

\end{enumerate}
Since the iteration is convergent, in this way, we obtain the optimal bandwidth $h$ that is completely data-based.

In the next section, we check whether or not our approach can work properly.

\begin{figure*}
\centerline{\includegraphics[width=1.90\columnwidth]{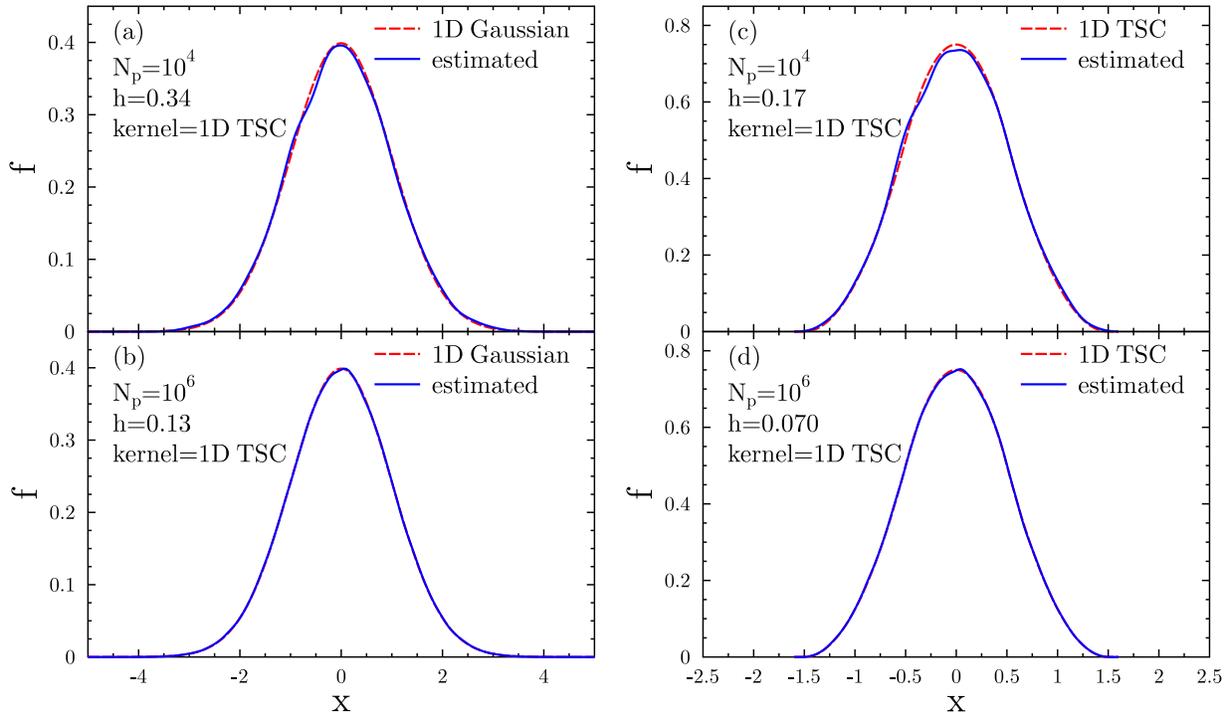}}
\caption{The estimated density functions are compared with analytic ones. Panels (a) and (b) are for 1D Gaussian function, and (c) and (d) are for 1D TSC function, respectively. The sample size $N_p$, the optimal bandwidth $h$ derived from our approach, are shown in the four panels. In all cases, we use 1D TSC as kernel function for the density estimation.}
\label{fig:hn}
\end{figure*}

\section{Results and Application}
\label{sec:resapp}

\subsection{Results for univariate samples with unimodal density}
\label{sec:reslt1}

As aforementioned, we need some experimental samples to verify our approach. For this purpose, we consider two sets of experimental data: (1) The first set of experimental data are drawn from the 1D standard normal distribution,
\begin{equation}
\label{eq:1dgauss}
f(x) = \frac{1} {\sqrt{2\pi}} e^{-\frac{x^2}{2}}.
\end{equation}
(2) The 1D TSC function can also be treated as a density function, and we draw the second set of samples from it by the usual {\it acceptance-rejection} method \cite{press92}.

With our approach, it takes just several iterations to obtain the optimal bandwidth $h$. In Figs.~\ref{fig:hopt-1} and \ref{fig:hopt-2}, we compare our results, which are purely data-based, with the analytic bandwidth $h$ of Eq.~(\ref{eq:hopt1}).  We can see that results obtained with 1D NGP as kernel function are slightly worse for small sample size $N_p$, but those based on 1D TSC functions give the best results, with the relative differences being only $ 2 \sim 3\%$. In Fig.~\ref{fig:hn}, we show the density functions estimated by using our optimal bandwidth $h$, to compare with the analytic density functions. We can see that the agreements between the estimated and the analytic densities are satisfactory.

\begin{figure}
\centerline{\includegraphics[width=1.0\columnwidth]{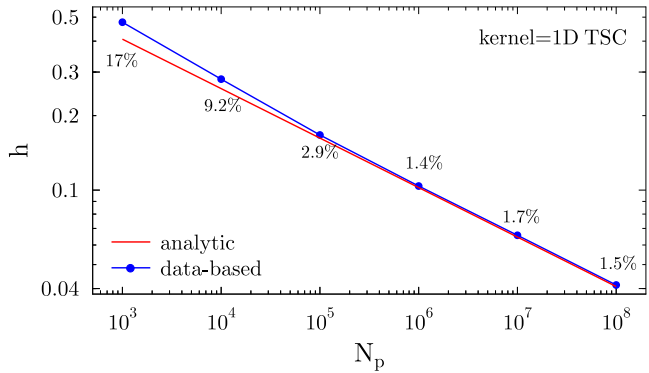}}
\caption{The same as Figs.~\ref{fig:hopt-1} and \ref{fig:hopt-2}, but the density function for the experimental sample is the multimodal density of Eq.~(\ref{eq:1dgaus3p}). We use 1D TSC as kernel function for the density estimation.}
\label{fig:hopt-mp}
\end{figure}

\begin{figure}
\centerline{\includegraphics[width=1.0\columnwidth]{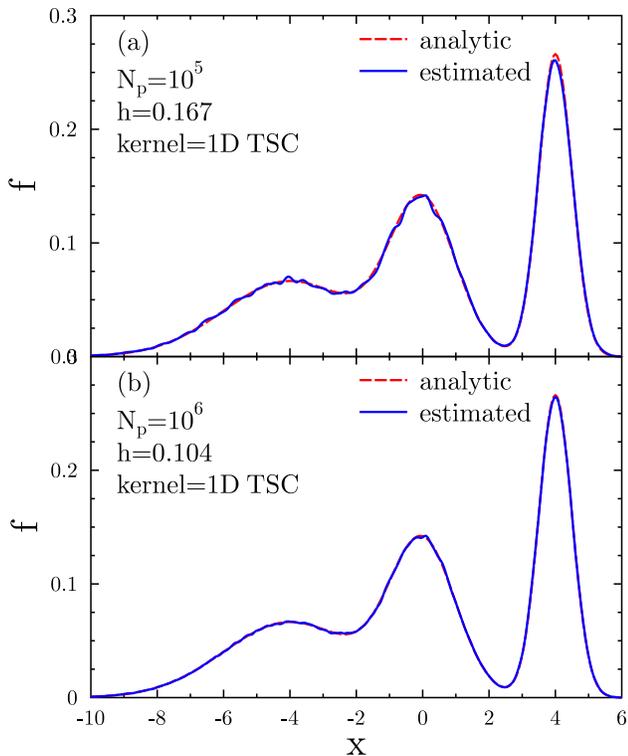}}
\caption{The estimated density functions are compared with analytic ones for multimodal density as Eq.~(\ref{eq:1dgaus3p}). The sample size $N_p$, the optimal bandwidth $h$ derived from our approach, are shown in the figure. We use 1D TSC as kernel function for the density estimation.}
\label{fig:hmp}
\end{figure}

\subsection{Results for univariate samples with multimodal density}
\label{sec:reslt2}

The approach is not restricted to just unimodal density, but can be also applied to cases of multimodal density functions. To check its validity for multimodal densities, we construct the test density function as the following,
\begin{equation}
\label{eq:1dgaus3p}
f(x) = \frac{1} {\sqrt{2\pi}} e^{-\frac{x^2}{2}} + \frac{1} {\sqrt{2\pi}\sigma_1} e^{-\frac{(x-x_1)^2}{2\sigma^2_1}} + \frac{1} {\sqrt{2\pi}\sigma_2} e^{-\frac{(x-x_2)^2}{2\sigma^2_2}},
\end{equation}
with $\sigma_1=2$, $x_1=-4$, $\sigma_2=0.5$, and $x_2=4$, respectively. In Fig.~\ref{fig:hopt-mp}, we compare our data-based bandwidth with the analytic results, and in Fig.~\ref{fig:hmp}, we compare the density functions estimated by our optimal bandwidth $h$ with the analytic density of Eq.~(\ref{eq:1dgaus3p}). It can be seen that the method is still be valid for estimation of multimodal densities.

\begin{figure}
\centerline{\includegraphics[width=1.0\columnwidth]{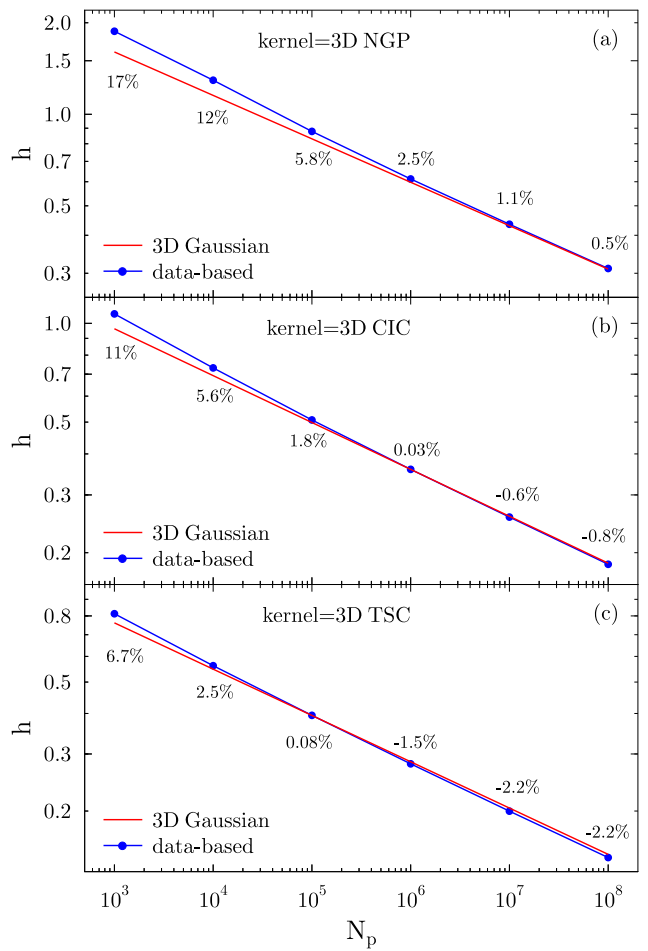}}
\caption{The same as Figs.~\ref{fig:hopt-1} and ~\ref{fig:hopt-2}, but the density function for the experimental sample is the 3D Gaussian function of Eq.~(\ref{eq:3dgauss}). Panels (a), (b) and (c) are for kernel functions of 3D NGP, 3D CIC and 3D TSC, respectively.}
\label{fig:hopt-3}
\end{figure}

\subsection{Results for three-dimensional samples}
\label{sec:reslt3}

With $|x|$ replaced by $r = (x^2_1 + x^2_2 + x^2_3)^{1/2}$, and with proper normalization, the 3D generalizations of the 1D kernel functions $W_{\rm NGP}(x)$, $W_{\rm CIC}(x)$, $W_{\rm TSC}(x)$ of Eqs.~(\ref{eq:ngp})-(\ref{eq:tsc}) are as follows
\begin{eqnarray}
\label{eq:kw3}
W_{\rm NGP3}(\vc{x}) & = & \frac{6}{\pi}W_{\rm NGP}(r), \nonumber \\
W_{\rm CIC3}(\vc{x}) & = & \frac{3}{\pi}W_{\rm CIC}(r), \nonumber \\
W_{\rm TSC3}(\vc{x}) & = & \frac{2}{\pi}W_{\rm TSC}(r),
\end{eqnarray}
in which $\vc{x}=(x_1, x_2, x_3)$. All these 3D kernel functions should satisfy
\begin{equation}
\label{eq:kernel3}
\int K_3(\vc{x})\dd \vc{x} = 1; {\hskip 20pt}  K_3(\vc{x})=K_3(-\vc{x}).
\end{equation}
Hence the estimated $\hat{f}_h(\vc{x})$ is
\begin{equation}
\label{eq:funkh3a}
\hat{f}_h(\vc{x}) = \frac{1}{N_p h^3}\sum\limits_{i=1}^{N_p}K_3(\frac{\vc{x}-\vc{x}_i}{h}).
\end{equation}

Eq.~(\ref{eq:hopt1}) is just the optimal bandwidth for univariate samples. Its 3D generalization is
\begin{equation}
\label{eq:hopt3}
h_{\rm opt}=\left[\frac{3 R_3(K_3)}{R_3(\nabla^2f)(\int x_1^2 K_3(\vc{x})\dd \vc{x})^2} \right]^{1/7} N_p^{-1/7},
\end{equation}
in which $K_3=K_3(\vc{x})$, $f=f(\vc{x})$, $\nabla^2=\partial^2_{x_1} + \partial^2_{x_2} + \partial^2_{x_3}$, and $R_3$ is defined as
\begin{equation}
\label{eq:rg3}
R_3(g)\equiv\int g(\vc{x})^2\dd \vc{x}.
\end{equation}
About Eq.~(\ref{eq:hopt3}), we refer the reader to Section-4.3.1 of Ref.~\cite{silverman86} for further details.

For a bandwidth $h$, the roughness $R_3(\nabla^2 f)$ in Eq.~(\ref{eq:hopt3}) is approximated by
\begin{equation}
\label{eq:chopt3}
R_3(\nabla^2 f) \approx R_3(\nabla^2\hat{f}_h) - \frac{42}{w^3 h^7 N_p}.
\end{equation}
Please see the explanation about Eq.~(\ref{eq:chopt3}) in Appendix-\ref{sec:apda}.

Proceed in the similar way as in Section~\ref{sec:method}, we use 3D standard normal distribution
\begin{equation}
\label{eq:3dgauss}
f(\vc{x}) = \frac{1}{(2\pi)^{3/2}} e^{-\frac{\vc{x}^2}{2}}
\end{equation}
as the experimental density function, from which we draw a random sample with size $N_p$, and with Eqs.~(\ref{eq:hopt3}) and (\ref{eq:chopt3}), we iterate to pick out the optimal bandwidth $h$.

In Fig.~\ref{fig:hopt-3}, we show our results to compare with the analytic bandwidth $h$ of Eq.~(\ref{eq:hopt3}). It can be seen that results obtained with 3D NGP as kernel function are obviously worse for small sample size $N_p$, with relative errors being 17\% and 12\% for $N_p=10^3$ and $10^4$ respectively. While computations based on 3D TSC functions give satisfactory results, with the relative differences being only $2 \sim 3\%$ for $N_p > 10^4$.

\subsection{A Cosmological Application}
\label{sec:app}

We apply our approach to a realistic model, the Hernquist model, to further verify its applicability. Hernquist \cite{hernquist90} proposed an analytic model of density profile for spherical galaxies and bulges, as
\begin{equation}
\label{eq:hernquist}
\rho(r)=\frac{M_{\rm T}}{2\pi}\frac{r_c}{r}\frac{1}{(r_c + r)^3},
\end{equation}
where $M_{\rm T}$ is the total mass, and $r_c$ is a scale length. With this density profile, we can generate the spatial distribution of a set of experimental sample. The code we used for generating this distribution is the initial-condition generator, taken from the $N$-body simulation code of Ref.~\cite{londrillo03}. In a previous work, we have shown that the code works very well when $r/r_c>0.05$ \cite{hep17}.

\begin{figure}
\centerline{\includegraphics[width=1.0\columnwidth]{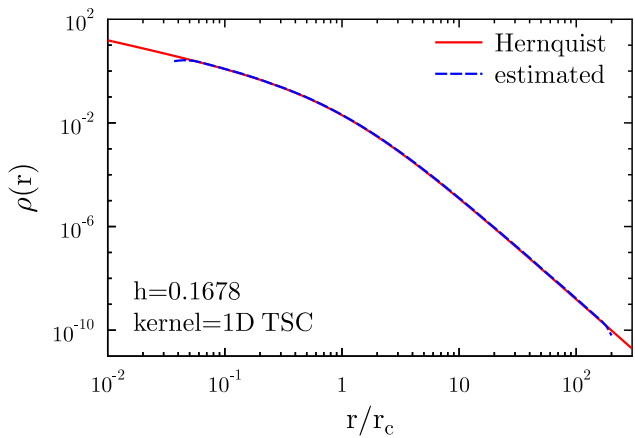}}
\caption{The estimated density profile is compared with the analytic one, i.e. the Hernquist \cite{hernquist90} model. The optimal bandwidth $h$ derived from our approach, is shown in the figure. The total mass in Eq.~(\ref{eq:hernquist}) of Hernquist model is set to be $M_{\rm T}=1$. The sample is generated with the initial-condition generator, taken from the $N$-body simulation code of \cite{londrillo03}, with the sample size $N_p=1.05 \times 10^6$. We use 1D TSC as kernel function for the density estimation.}
\label{fig:rhor}
\end{figure}

Fig.~\ref{fig:rhor} compares the estimated density profile by the 1D kernel estimation approach with the analytic model. For Hernquist model, the analytic optimal bandwidth given by Eq.~(\ref{eq:hopt1}) is $h=0.1712$, while our approach gives $h=0.1678$, with only -1.9\% relative error from the analytic one.

It can be seen that our approach gives very satisfactory result.

\section{Summary and Conclusions}
\label{sec:summy}

In many scientific disciplines such as statistics and modern cosmology, it is a common practice to estimate the probability density function or matter density function in a given spatial region from statistical, observational or simulation samples. Kernel density estimation is widely used among all the approaches, but to select an optimal bandwidth of kernel estimation, which is completely based on data samples, is a long-term issue that has not been well settled so far. The formulae of Eqs.~(\ref{eq:hopt1}) and ~(\ref{eq:hopt3}) do provide the analytic optimal bandwidth, but they cannot be applied directly to data samples, since they depend on the unknown underlying density functions from which the samples are drawn.

In this work, we devise an approach to pick out the totally data-based optimal bandwidth. First, we derive the correction formulae, Eq.~(\ref{eq:chopt1}) for 1D or Eq.~(\ref{eq:chopt3}) for 3D samples, to compute the roughness of the sample's density function. Then substitute the correction formulae into the analytic formulae Eqs.~(\ref{eq:hopt1}) or (\ref{eq:hopt3}) for optimal bandwidth, and through iteration we obtain the sample's optimal bandwidth. The whole process terminates just after several iterations, and compared with the analytic formulae, our approach gives very good results, whose relative differences from the analytic formulae are only $2 \sim 3\%$ for sample size larger than $10^4$.

Technically, we use NGP, CIC and TSC mass assignment schemes as kernel functions for density estimation, and use the Gaussian and TSC functions as well as Ref.~\cite{hernquist90}'s model as experimental density functions, but we emphasis that the results should not depend on the adopted kernel functions or density functions.

Incidentally, if we divide the whole sample into several parts, and apply our approach of selecting optimal bandwidth to sub-samples part by part, then our approach is not hard to be generalized to cases of variable kernel estimations.

\acknowledgments
\label{ack}

This work is supported by the National Science Foundation of China (no.~11273013), and also supported by the Natural Science Foundation of Jilin Province, China (no.~20180101228JC).

\appendix
\section {Correction Formula for the Roughness of Density Functions}
\label{sec:apda}

We give the derivation of the correction formula in Eq.~(\ref{eq:chopt1}) for the roughness of a 1D density function.

The estimated density $\hat{f}_h(x)$ of the sample is a random variable for any $x$, and should obey Poisson distribution. $\hat{f}_h(x)$ can be written as
\begin{equation}
\label{eq:diff}
\hat{f}_h(x) =  f(x) + \Delta\hat{f}_h(x),
\end{equation}
where $f$ is the unknown underlying density function, from which the sample is drawn, and $\Delta\hat{f}_h$ is the difference between the two, whose ensemble average is of course $\overline{\Delta\hat{f}_h} = 0$.

We require both $\hat{f}_h$ and $f$ satisfy the normalization $\int \hat{f}_h(x) dx = \int f(x) dx = 1$. So we have $\hat{f}_h(x_i) \Delta x =  {n_i}/{N_p}$, in which $n_i$ is the counted particle number at $x_i$ within the interval $\Delta x$. We roughly estimate $\Delta x \approx w h$, in which $w$ is the width of the kernel function. From Fig.~\ref{fig:wx}, we see that $w=1$, $2$ and $3$ for NGP, CIC and TSC, respectively. Hence, $\hat{f}_h(x_i)$ is estimated as
\begin{equation}
\label{eq:fhiw}
\hat{f}_h(x_i) \approx \frac{n_i}{w h N_p},
\end{equation}
and hence
\begin{equation}
\label{eq:afhiw}
f(x_i) = \overline{\hat{f}_h(x_i)} \approx \frac{\overline{n_i}}{w h N_p}.
\end{equation}

Calculating the second derivative and taking the square of both sides of Eq.~(\ref{eq:diff}), we have
\begin{equation}
\label{eq:diff2}
\hat{f}''_h(x)^2 =  f''(x)^2 + 2f''(x)\Delta\hat{f}''_h(x) + \Delta\hat{f}''_h(x)^2,
\end{equation}
in which the second derivative of the estimated density $\hat{f}''_h$ is evaluated by
\begin{equation}
\label{eq:dev2nd}
\hat{f}''_h(x_i) = \frac{ \hat{f}_h(x_{i+1}) + \hat{f}_h(x_{i-1}) -2\hat{f}_h(x_i)} {h^2},
\end{equation}
in which $x_i = x_{i+1} - h = x_{i-1} + h$. Making the ensemble average of $\hat{f}''_h(x)^2$, we obtain
\begin{eqnarray}
\label{eq:diff3}
\overline{\hat{f}''_h(x)^2} & = & \overline{f''(x)^2} + 2\overline{f''(x)\Delta\hat{f}''_h(x)} + \overline{\Delta\hat{f}''_h(x)^2} \nonumber \\
 & = & f''(x)^2 + \overline{\Delta\hat{f}''_h(x)^2},
\end{eqnarray}
in which $f''(x)$ is not random variable, and $\overline{\Delta\hat{f}''_h(x)} = 0$. Further, from Eq.~(\ref{eq:dev2nd}), we have
\begin{eqnarray}
\label{eq:diff4}
\overline{\Delta\hat{f}''_h(x_i)^2} & = & \frac{1}{h^4}( \overline{\Delta\hat{f}_h(x_{i+1})^2} + \overline{\Delta\hat{f}_h(x_{i-1})^2} + 4 \overline{\Delta\hat{f}_h(x_i)^2}  \nonumber \\
 & & + 2\overline{\Delta\hat{f}_h(x_{i+1})\Delta\hat{f}_h(x_{i-1})} - 4 \overline{\Delta\hat{f}_h(x_{i+1})\Delta\hat{f}_h(x_i)} \nonumber \\
 & & - 4 \overline{\Delta\hat{f}_h(x_i)\Delta\hat{f}_h(x_{i-1})}) = \frac{1}{h^4}( \overline{\Delta\hat{f}_h(x_{i+1})^2} \nonumber \\
 & & + \overline{\Delta\hat{f}_h(x_{i-1})^2} + 4 \overline{\Delta\hat{f}_h(x_i)^2}).
\end{eqnarray}
Since the deviation $\Delta \hat{f}_h$'s are independent at different $x$, to arrive at the last equality in Eq.~(\ref{eq:diff4}), we use
\begin{eqnarray}
\overline{\Delta\hat{f}_h(x_{i+1})\Delta\hat{f}_h(x_{i-1})} & \approx & \overline{\Delta\hat{f}_h(x_{i+1})} \cdot\overline{\Delta\hat{f}_h(x_{i-1})} = 0,  \nonumber \\ \overline{\Delta\hat{f}_h(x_{i+1})\Delta\hat{f}_h(x_i)} & \approx & \overline{\Delta\hat{f}_h(x_{i+1})} \cdot\overline{\Delta\hat{f}_h(x_i)} = 0,  \nonumber \\ \overline{\Delta\hat{f}_h(x_i)\Delta\hat{f}_h(x_{i-1})} & \approx & \overline{\Delta\hat{f}_h(x_i)} \cdot\overline{\Delta\hat{f}_h(x_{i-1})} = 0, \nonumber
\end{eqnarray}
Also, since $n(x) = N_p \hat{f}_h(x) w h$ satisfies the Poisson distribution at $x$, so
\begin{eqnarray}
\overline{\Delta\hat{f}_h(x_{i+1})^2} & = & \frac{\overline{n_{i+1}}}{w^2 h^2 N^2_p} \approx \frac{f(x_{i+1})}{w h N_p}, \nonumber \\
\overline{\Delta\hat{f}_h(x_i)^2} & = & \frac{\overline{n_i}}{w^2 h^2 N^2_p} \approx \frac{f(x_i)}{w h N_p}, \nonumber \\
\overline{\Delta\hat{f}_h(x_{i-1})^2} & = & \frac{\overline{n_{i-1}}}{w^2 h^2 N^2_p} \approx \frac{f(x_{i-1})}{w h N_p}, \nonumber
\end{eqnarray}
in which we considered Eqs.~(\ref{eq:fhiw}) and (\ref{eq:afhiw}). So Eq.~(\ref{eq:diff4}) is simplified as
\begin{equation}
\label{eq:diff5}
\overline{\Delta\hat{f}''_h(x_i)^2} \approx  \frac{1}{w h^5 N_p}(f(x_{i+1}) + 4 f(x_i) + f(x_{i-1})).
\end{equation}
Substitute Eq.~(\ref{eq:diff5}) into Eq.~(\ref{eq:diff3}), and integrate both sides, we obtain
\begin{eqnarray}
\label{eq:diff6}
\int\overline{\hat{f}''_h(x)^2} \dd x & = & \int f''(x)^2 \dd x + \int \overline{\Delta\hat{f}''_h(x)^2} \dd x, \nonumber \\
 & \approx & \int f''(x)^2 \dd x + \frac{6}{w h^5 N_p}.
\end{eqnarray}
Since the ensemble average $\overline{\hat{f}''_h(x)^2}$ is not easy to derive, in practice we can simply approximate it as $\hat{f}''_h(x)^2$, and finally we obtain Eq.~(\ref{eq:chopt1}).

It's not hard to generalize the one-dimensional formula Eq.~(\ref{eq:chopt1}) to 3D case. For 3D samples, Eq.~(\ref{eq:fhiw}) should be adjusted as
\begin{equation}
\label{eq:fhiw3}
\hat{f}_h(\vc{x}_i) \approx \frac{n_i}{w^3 h^3 N_p}.
\end{equation}
In the expression of $R_3(\nabla^2\hat{f}_h)$, there are total 3 squared terms of the second derivative $(\partial^2 \hat{f}_h/\partial x^2_\alpha)^2$ ($\alpha=1,2,3$), and each contributes a correction $6/w^3 h^7 N_p$. Also, there are total 6 mixed products of the second derivative as $(\partial^2 \hat{f}_h/\partial x^2_\alpha) (\partial^2 \hat{f}_h/\partial x^2_\beta)$ ($\alpha, \beta=1,2,3, \alpha\neq\beta$), and each contributes a correction $4/w^3 h^7 N_p$. So finally, these corrections are added up to $42/w^3 h^7 N_p$, and we obtain Eq.~(\ref{eq:chopt3}).



\begin{thebibliography}{99}
\bibitem{efs85} G. Efstathiou, M. Davis, C. S. Frenk, and S. D. M. White, \apj {\bf 57} (1985) 241.

\bibitem{bert98} E. Bertschinger, \araa {\bf 36} (1998) 599.

\bibitem{Springel01} V. Springel, N. Yoshida, and S. D. M. White, \na {\bf 6} (2001) 79.

\bibitem{nfw95} J. F. Navarro, C. S. Frenk, and S. D. M. White, \mnras {\bf 275} (1995) 720.

\bibitem{nfw96} J. F. Navarro, C. S. Frenk, and S. D. M. White, \apj {\bf 462} (1996) 563.

\bibitem{nfw97} J. F. Navarro, C. S. Frenk, and S. D. M. White, \apj {\bf 490} (1997) 493.

\bibitem{moore99} B. Moore, T. Quinn, F. Governato, J. Stadel, and G. Lake, \mnras {\bf 310} (1999) 1147.

\bibitem{einasto65} J. Einasto, Trudy Inst. Astrofizicheskogo Alma-Ata {\bf 51} (1965) 87.

\bibitem{navarro10} J. F. Navarro, A. D. Ludlow, V. Springel, J. Wang, M. Vogelsberger, S. D. M. White, A. Jenkins, C. S. Frenk, and A. Helmi, \mnras {\bf 402} (2010) 21.

\bibitem{Schaap2000} W. E. Schaap and R. van de Weygaert, \aap {\bf 363} (2000) L29.

\bibitem{Abel2012} T. Abel, O. Hahn, and R. Kaehler, \mnras {\bf 427} (2012) 61.

\bibitem{Shandarin2012} S. Shandarin, S. Habib, and K. Heitmann, \prd {\bf 85} (2012) 083005.

\bibitem{silverman86} B. W. Silverman, {\it Density Estimation for Statistics and Data Analysis}, Chapman and Hall, London (1986).

\bibitem{scott92} D. W. Scott, {\it Multivariate Density Estimation: Theory, Practice, and Visualization}, Wiley, New York (1992).

\bibitem{gentle09} J. E. Gentle, {\it Computational Statistics}. Springer Science+Business Media, New York (2009).

\bibitem{jones96} M. C. Jones, J. S. Marron, and S. J. Sheather, J. Am. Stat. Assoc. {\bf 91} (1996) 401.

\bibitem{wolpert13} D. H. Wolpert and S. DeDeo, Entropy {\bf 15} (2013) 4668.

\bibitem{hep14} N. Sui, M. Li, and P. He, \mnras {\bf 445} (2014) 4211.

\bibitem{Rudemo1982} M. Rudemo, Scand. J. Statist. {\bf 9} (1982) 65.

\bibitem{Bowman1984} A. W. Bowman, Biometrika {\bf 71} (1984) 353.

\bibitem{Hall1992} P. Hall, J. S. Marron, and B. U. Park, Probab. Theory Relat. Fields {\bf 92} (1992) 1.

\bibitem{Sheather1991} S. J. Sheather and M. C. Jones, J. R. Statist. Soc. B {\bf 53} (1991) 683.

\bibitem{Botev2010} Z. I. Botev, J. F. Grotowski, and D. P. Kroese, Annals of Statistics {\bf 38} (2010) 2916.

\bibitem{press92} W. H. Press, S. A. Teukolsky, W. T. Vetterling, and B. P. Flannery, {\it Numerical Recipes in Fortran - The Art of Scientific Computing}, 2nd edt., Cambridge University Press (1992).

\bibitem{hockney81} R. W. Hockney and J. W. Eastwood, {\it Computer Simulations Using Particles}, McGraw-Hill, New York (1981).

\bibitem{hernquist90} L. Hernquist, \apj {\bf 356} (1990) 359.

\bibitem{londrillo03} P. Londrillo, C. Nipoti, and L. Ciotti, Mem. Soc. Astron. Ital. {\bf 1} (2003) 18.

\bibitem{hep17} N. Sui and P. He, Int. J. Mod. Phys. D {\bf 26} (2017) 1750130.

\end{thebibliography}
\end{document}